\documentclass[aps,prd,twocolumn,groupedaddress]{revtex4}  
\usepackage{graphicx}
\usepackage{booktabs}
\usepackage{amssymb,bm,mathrsfs,bbm,amscd}
\usepackage[tbtags]{amsmath}
\usepackage{epstopdf}
\begin{document}
\title{Sensitivity functions for space-borne gravitational wave detectors}
\author{Xiao-Yu Lu}
\author{Yu-Jie Tan}\email[E-mail:]{yjtan@hust.edu.cn}
\author{Cheng-Gang Shao}

\affiliation{MOE Key Laboratory of Fundamental Physical Quantities Measurement and Hubei\\ Key Laboratory of Gravitation and Quantum Physics, PGMF and School of Physics, Huazhong University of Science and Technology, Wuhan 430074, P. R. China}

\date{\today}

\begin{abstract}
Time-delay interferometry is put forward to improve the signal-to-noise ratio of space-borne gravitational wave detectors by canceling the large laser phase noise with different combinations of measured data. Based on the Michelson data combination, the sensitivity function of the detector can be obtained by averaging the all-sky wave source positions. At present, there are two main methods to encode gravitational wave signal into detector. One is to adapt gravitational wave polarization angle depending on the arm orientation in the gravitational wave frame, and the other is to divide the gravitational wave signal into plus and cross polarizations in the detector frame. Although there are some attempts using the first method to provide the analytical expression of sensitivity function, only a semianalytical one could be obtained. Here, starting with the second method, we demonstrate the equivalence of both methods. First time to obtain the full analytical expression of sensitivity function, which provides a fast and accurate mean to evaluate and compare the performance of different space-borne detectors, such as LISA and TianQin.
\end{abstract}

\maketitle

\section{INTRODUCTION}\label{section1}

One hundred years ago, Einstein predicted the gravitational wave (GW) in general relativity (GR), which propagates oscillations of the gravitational field in spacetime. Since this oscillating signal carries information about physics and dynamics of the wave sources, which helps to test GR and observe the evolution of universe, studying how to detect GW is significantly necessary, and a lot of related researches have been performed \cite{1b,1c,1d,1e,1f,1g,1,2,3,2a,2b,2c,2d,3a,8,3b,9,3d}, such as LIGO/VIRGO \cite{2a,2b,2c,2d}, DECIGO \cite{3a}, LISA \cite{8}, ASTROD-GW \cite{3b}, OMEGA \cite{1d}, TianQin \cite{9}, TAIJI \cite{3d}, etc. As a ground-based detector, the GW signal has first successfully detected by advanced LIGO \cite{3}, and then more GW events have been observed \cite{4,5,6,7}, which opens up a new era of GW astronomy. Complementary to ground-based interferometers sensitive to high-frequency band signal, LISA is put forward to test the low-frequency band signal, which is the most notable example of space-based detectors. TianQin space mission has been also proposed to detect this similar frequency domain signal, which aims at observing the GW emitted by a special source: J0806 \cite{9,10,10a}.

To determine whether GW signal is detectable or not for a given detector, it is extremely necessary to know its sensitivity limit \cite{12,13,14,15,16}, which depends on the amplitudes of signal and noise in the output of the measured instrument. To improve the detectable strength, one should make a lot of efforts to reduce the effect of different noise processes. Usually, laser phase noise is rather large in the detecting system. For the ground-based interferometers, two arms are fixed with equal lengths to make the lasers experience identical delay; therefore, the laser phase noise can be canceled very well. However, it is impossible to maintain the length of each arm constant for the space-based interferometers. This results in that the lasers in different arms have different delays and residual laser phase noise greatly affects the detecting sensitivity. To solve this problem, Tinto \emph{et al.} first proposed time-delay interferometry (TDI) technique \cite{17} to cancel this noise with different combinations of measured data, in which the average over the source directions and polarization states has been done via Monte Carlo computer simulation. In addition, some other explorations for developing this technique have also been done \cite{17a,16a,18,19,20,21,22,23,24,25,26,27,28}. However, most of the above works just showed the mathematical simulation. In this paper, we focus on giving a full analytical expression of the transfer function with a Michelson data combination.

As a GW response of space-borne detection is dependent on the information of sources and orientations of the detectors, it is reasonable to make the inclination, polarization, and sky average and finally obtain an analytical expression of transfer function. In the analysis of GW response, the signal is encoded through two methods. The first method is to adopt the GW polarization angle dependent on the arm orientations of the detector, so the projections of GW signal on the two interference arms are different in the GW frame \cite{12}. The second method is to divide the signal into two polarization components in the detector frame \cite{38,39}. Although there are lots of discussions on these two methods \cite{36,37}, there still lacks the discussion on analytical formulas. Some previous studies with the first method have tried to provide the analytical expression of sensitivity function \cite{12,30a,35a}, but the result is like a kind of semianalytical one. Although the second method is mainly used for numerical simulation, we here start from this method to calculate the contribution of plus and cross polarizations for both interference arms, compare the result of the transfer function with that of the first method, and finally demonstrate the equivalence of these two methods. Furthermore, we first obtain a full analytical expression of all-sky averaged sensitivity function with the second method, which is useful to compare the performances of different space-borne detectors, such as LISA and TianQin.

The paper is organized as follows. In Sec. II, based on the Michelson data combination, we review the application of TDI technique on canceling laser phase noise. In Sec. III, we demonstrate the equivalence of two methods to obtain the all-sky averaged transfer function. In Sec. IV, combining the noise and signal analysis, we give the analytical expression of the sensitivity function, and further apply it on the cases of LISA and TianQin missions. The paper is concluded in Sec. V.

\section{Time-delay interferometry and noise cancellation}\label{section2}

Space-borne GW detection essentially makes a laser beam propagating from a remote spacecraft (SC) beat with a reference beam in local SC. GW in the spacetime influences the light path and further the beating signal. Thus, to detect the GW signal is to measure the Doppler shifts of the laser frequency. Analogous to the work of Estabrook \emph{et.al} \cite{19} on TDI technique for LISA mission, we try to give a detailed analytic study on the applications of TDI technique for space-borne detectors, and the noise analysis is focused on in this section.

\begin{figure}[!t]
\includegraphics[width=0.4\textwidth]{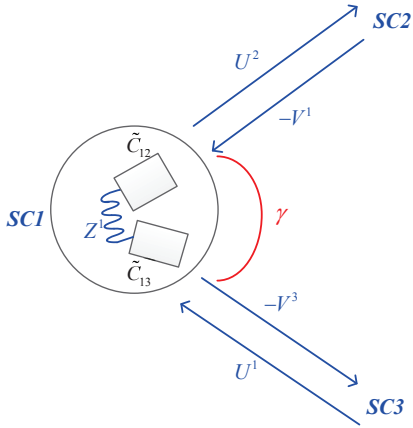}
\caption{\label{fig1} Triangle configuration for space-borne gravitational wave detector. Here, we have adopted the same convention as that in \cite{25}.}
\end{figure}

The geometry of the space-borne detector with the laser beams is shown in Fig. 1: three spacecrafts (SC1, SC2, SC3) fly in a triangle formation, where the angle $\gamma $ of SC2 and SC3 with respect to SC1 is arbitrary; every SC contains two proof masses and two lasers, respectively, mounted on the optical benches, and ${Z^i}$ represents the fractional frequency fluctuation of the two laser beams exchanged between the optical benches; three SCs are joined by laser beams, and the time delay operation $E_{ij}$ of the beam from $i$th SC to $j$th one is defined as $E_{ij}f(t)\equiv f(t-L_{ij}/c)$ with $L_{ij}$ being the arm length and $c$ being the speed of light. $\tilde{C}_{ij}$ represents the noises of laser phase and optical benches in the $i$th SC facing the $j$th one. ${U^i}$ and ${V^i}$ ($i = 1,2,3$) represent the fractional frequency fluctuations for the beams propagating to the $i$th SC. For convenience, we only consider laser phase, proof mass and shot noises. Data streams for SC1 can be expressed as \cite{25}
\begin{eqnarray}\label{n1}
&&{U^1} = {E_{31}}{{\tilde C}_{31}} - {{\tilde C}_{13}} + 2{\delta_{13}} + {h_{U^1}} + Y_{U^1}^{\rm{shot}}, \nonumber\\
&&{V^1} =  - {E_{21}}{{\tilde C}_{21}} + {{\tilde C}_{12}} + 2{\delta_{12}} - {h_{V^1}} - Y_{V^1}^{\rm{shot}}, \nonumber\\
&&{Z^1} = {{\tilde C}_{12}}-{{\tilde C}_{13}} + {\delta_{12}} + {\delta_{13}}.
\end{eqnarray}
Here, $\delta _{ij}$ is the fluctuation induced by the random velocity noise of the proof mass, $h_{U^1}$ and $h_{V^1}$ represent GW signal, and $Y^{\rm{shot}}$ is the fluctuation due to shot noise. For the noise-canceling combination ${ \sum\limits_{i = 1}^3 {{p_i}{V^i} + {q_i}{U^i} + {r_i}{Z^i}}}$, power spectral densities (PSDs) of proof mass and shot noises \cite{25} can be expressed as
\begin{eqnarray}\label{n2}
&&{N_{\rm{pf}}} =( {\sum\limits_{i = 1}^3 {\left| {2{p_i} + {r_i}} \right|{}^2 + \left| {2{q_i} + {r_i}} \right|{}^2} } ){S_{{\rm{pf}}}}, \nonumber \\
&&{N_{\rm{shot}}} = ( {\sum\limits_{i = 1}^3 {\left| {{p_i}} \right|{}^2 + \left| {{q_i}} \right|{}^2} } ){S_{{\rm{shot}}}}, \hfill
\end{eqnarray}
with ${S_{\rm{pf}}} \equiv \frac{{s_a^2}}{{{{\left( {2\pi fc} \right)}^2}}}$ and ${S_{\rm{shot}}} \equiv \frac{{{{\left( {2\pi f} \right)}^2}s_{x}^2}}{{{c^2}}}$, where ${s_a}$ and ${s_x}$ are amplitude spectral densities (ASDs) of proof mass acceleration and shot noises, respectively.

For the GW detection, different data combinations may produce different link configurations. The more links are, the more experimental data are utilized, and the better sensitivities will be. However, to derive the analytic expression of sensitivity function, one usually starts with the four-link Michelson data combination, since the calculation is not that complex. For the Michelson data combination
\begin{eqnarray}\label{n3}
X\!\!\!\!\!\!&&\equiv \begin{bmatrix} {p_1,p_2,p_3} \\ {q_1,q_2,q_3} \\ {r_1,r_2,r_3} \\ \end{bmatrix}\nonumber \\
  &&=\begin{bmatrix} {1 - {E_{13}}{E_{31}},\;\;0,\;\;{E_{31}}({E_{12}}{E_{21}} - 1)} \\{1 - {E_{12}}{E_{21}},\;\;{E_{21}}({E_{13}}{E_{31}} - 1),\;\;0 \;\;} \\ {({E_{13}}{E_{31}} - 1)(1 - {E_{12}}{E_{21}}),\;\;0,\;\;0\;\;} \\ \end{bmatrix},
\end{eqnarray}
the calculation can be further simplified in frequency domain by assuming $E_{ij} = {e^{i\Omega L/c}}$, i.e., all of the arm lengths are assumed to be equal as $L$. As this data combination can cancel the noises of laser phase and optical benches, PSD for total noises of the Michelson combination can be finally given by \cite{16a}
\begin{eqnarray}\label{n4}
\!\!\!\!\!\!{S_N^{\!(\!X\!)\!}}(\!f) \!\!\!&&\!\!\!\!\!\equiv {N_{\rm{pf}}} + {N_{\rm{shot}}} \nonumber \\
&&\!\!\!\!\!=\!\!({8{{\sin }^2}\!\frac{{4\pi fL}}{c}\!\!+\!\!32{{\sin }^2}\!\frac{{2\pi fL}}{c}})\!S_{\rm{pf}}\!\!+\!\! 16{\sin ^2}\!\frac{{2\pi fL}}{c}\!\!S_{\rm{shot}}
\end{eqnarray}
with $f=\Omega/2\pi$ being the GW frequency.

\section{The transfer function of space-borne gravitational wave detectors}\label{section3}

Combining with the Michelson data combination, we discuss two methods on encoding GW signal into the detectors. One is to adopt the polarization angle dependent on the arm orientation in GW frame, which makes the projections of GW signal on the two interference arms different and is usually used for the four-link configurations; the other is to divide the GW signal into plus and cross polarizations in detector frame, which is usually used for more general combinations, such as four-link, five-link, and full six-link configurations.

\subsection{The first method related to two different polarization angles}\label{subsection3.1}

The responses of GW are different for the different interference arms due to the polarization angle dependent on the arm orientation in GW frame.

First, the single arm case is considered. Assume GW travels along the $z$ direction and the detection arm is in $x$-$z$ plane with an angle $\theta$ from $z$ axis. The laser beam is initially sent out by one SC, received by another SC, and then back to the first SC. Here, the trajectory of the laser beam is characterized by a null geodesic along
\begin{eqnarray}\label{m1}
0=d{s^2}\!\!=\!\!\!\!&&\!\!- {c^2}d{t^2}\!\!+\!\!d{z^2}\!\!+\!\!(1 + h\cos 2\psi )d{x^2}\!\!+\!\!(1\!\!-\!\!h\cos 2\psi )d{y^2}\nonumber\\
&&- h\sin 2\psi dxdy,
\end{eqnarray}
with $h(t-z/c)$ the GW amplitude and $\psi$ the polarization angle. Since the GW perturbation causes the frequency shift in the round-trip journey, the frequency shift can be obtained as \cite{31}
\begin{eqnarray}\label{m2}
z(t,\theta,\psi)\!\!=\!\!\frac{1}{2}\!\cos 2{\psi}[ {(1\!\!-\!\!{\mu})\!\!+\!\!2{\mu}{e^{ - iu(1+{\mu})}}\!\!\!\!-\!\!(1\!\!+\!\!{\mu}){e^{ - i2u}}} ]h(t),
\end{eqnarray}
with $\mu=\cos \theta$ and $u\equiv\frac{{2\pi fL}}{c}$.

\begin{figure}[!t]
\includegraphics[width=0.4\textwidth]{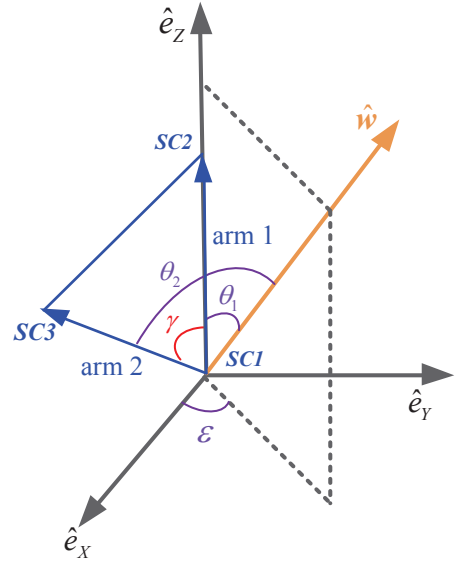}
\caption{The GW frame. The signal propagates along $\hat{w}$ direction, which has the angle $\theta_1$ and $\theta_2$ with the detector arm 1 and 2, respectively. The detector arms 1 and 2, respectively, locate in the $\hat{e}_X-\hat{e}_Z$ plane, and the angle between the two interference arms is $\gamma$.}
\label{fig7}
\end{figure}

For the detector with two interference arms, the GW frame is established in Fig.\,{\ref{fig7}}  \cite{34}. GW propagates along the $\hat{w}$ direction with angle $\theta_1$ from the $\hat{e}_X-\hat{e}_Z$ plane. The GW  propagation direction can be written as
\begin{align}\label{A1}
\hat{w}= \sin \theta_1 \cos \varepsilon \hat{e}_X + \sin \theta_1 \sin \varepsilon \hat{e}_Y + \cos \theta_1 \hat{e}_Z,
\end{align}
and the orthonormal basis vectors are
\begin{align}\label{A2}
&\hat{\theta_1} = \cos \theta_1 \cos \varepsilon \hat{e}_X+ \cos \theta_1 \sin \varepsilon \hat{e}_Y - \sin \theta_1 \hat{e}_Z, \nonumber\\
&\hat{\varepsilon}= - \sin\varepsilon \hat{e}_X + \cos \varepsilon \hat{e}_Y.
\end{align}
The position vectors of three SCs are
\begin{align}\label{A3}
&\vec{r}_1 = 0, \nonumber \\
&\vec{r}_2= L\hat{e}_X, \nonumber \\
&\vec{r}_3= L(\sin  \gamma\hat{e}_X+\cos \gamma\hat{e}_Y).
\end{align}
Assuming the detector arm between $SC1$ and $SC2$ is arm 1, and that between $SC1$ and $SC3$ is arm 2, shown in Fig. \ref{fig7}, the angles between the detection arms and GW propagation direction are, respectively, $\theta_1$ and $\theta_2$, the polarization angles are $\psi_1$ and $\psi_2$, and the detector signals in the two interference arms can be written as $z_1$ and $z_2$.

This method paves the way to calculate the approximately analytic expression of the transfer function for space-borne GW detectors. In general, since one does not know where GW comes from, the source positions and polarizations should be averaged, and then the transfer function can be obtained as \cite{12}
\begin{eqnarray}\label{m3}
\mathcal{R}(u) \equiv \frac{1}{{4\pi }}\int\limits_{ - \pi }^\pi  {d\varepsilon } \int\limits_0^\pi  {\sin {\theta _1}d{\theta _1}}  \cdot \frac{1}{{2\pi }}\int\limits_0^{2\pi } {d\psi_1 } {\left| {{z_1} + {z_2}} \right|^2}.
\end{eqnarray}
These references \cite{10,30a} have made a lot of efforts to obtain the analytic expression of Eq.\,({\ref{m3}}) as
\begin{eqnarray}\label{m4}
\mathcal{R}(u) = &&\!\!\!\!\!\!4\sin ^2u[ (1\!+\!\cos ^2\beta u)( {\frac{1}{3}\!\!-\!\!\frac{2}{{{u^2}}}} )\!+\!\sin^2u\!+\!\frac{4}{{{u^3}}}\sin u\cos u]\nonumber \hfill \\
&&\!\!\!\!\!\!-\!\!\frac{1}{\pi }\sin ^2u\int_0^{2\pi }\!\!d \varepsilon\!\!\int_{ - 1}^{ + 1} \!\!d {\mu _1}\left( {1\!-\!2{{\sin }^2}\alpha } \right)\!\eta \left( {u,{\theta _1},{\theta _2}} \right)
\end{eqnarray}
with
\begin{eqnarray}\label{m5}
\!\!\!\!\!\!\!\!\!\!\eta \left( {u,{\theta _1},{\theta _2}} \right)\!\!=&&\!\!\!\!\!\!\!( {\cos u\!-\!\cos  {u{\mu _1}}} )[\cos u\!-\! \cos (u{\mu _2})]{\mu _1}{\mu _2} \nonumber \\
&&\!\!\!\!\!\!\!+[ {\sin u\!\!-\!\!\mu _1 \sin (u\mu _1)}][ {\sin u\!\!-\!\!\mu _2 \sin (u\mu _2)}],
\end{eqnarray}
which is the present finest analytic expression for four-link configurations. Unfortunately, this expression still has the remaining integral term, which is not a thoroughly analytic expression. With the approximate formation of this result, many references \cite{35,35a,31,34,30a} have analyzed some other physical problems. To solve this semianalytic expression problem, we recalculate the sensitivity function with another method, which will be discussed in detail in the following sections.

\subsection{The second method related to plus and cross polarizations}\label{subsection3.2}

The transfer function can also be obtained from the viewpoint of the two different GW polarizations, and this method is usually used for different link configurations.

\begin{figure}[!t]
\includegraphics[width=0.5\textwidth]{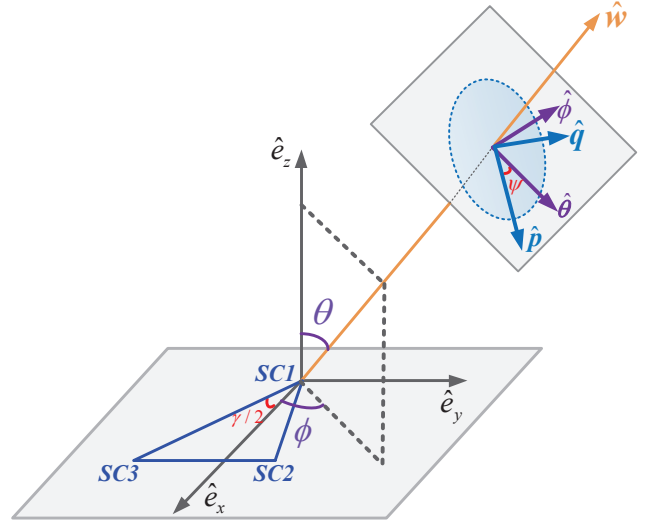}
\caption{\label{fig2} The detector frame \cite{29,30}.}
\end{figure}

We choose the detector frame $(\hat{e}_x,\hat{e}_y,\hat{e}_z)$ mounted on the center of SC1 (see Fig.\,\ref{fig2}): $\hat{e}_x$ is the direction pointing toward the middle point between SC2 and SC3, $\hat{e}_z$ the normal direction of the triangle detector plane, and $\hat{e}_y$ completes the reference frame. To understand the response of GW well, another two coordinate frames are introduced: the observational reference frame (ORF: $ \hat{\theta},\hat{\phi},\hat{w}$) and the canonical reference frame (CRF: $\hat{p},\hat{q},\hat{w}$). Here, $\hat{p}$, $\hat{q}$ represent the directions of the two polarization axes of the gravitation radiation, and $\hat{w}$ is the direction of GW source seen by an observer at rest in SC1. In ORF, $\hat{w}$ can be rewritten as \cite{16a,21,26}
\begin{eqnarray}\label{n5}
\hat{w}\equiv-\hat{k}= \sin \theta \cos \phi \hat{e}_x + \sin \theta \sin \phi \hat{e}_y + \cos \theta \hat{e}_z,
\end{eqnarray}
with $\hat{k}$ being the GW propagating direction, and the transverse plane is spanned by the unit transverse vector $\hat{\theta}$ and $\hat{\phi}$, which are defined as
\begin{eqnarray}\label{n6}
&&\hat{\theta} \equiv \partial \hat{w}/\partial\theta= \cos \theta \cos \phi \hat{e}_x + \cos \theta \sin \phi \hat{e}_y - \sin \theta \hat{e}_z, \nonumber\\
&&\hat{\phi}\equiv \partial\hat{w}/\left( {\sin \theta \partial\phi } \right)= - \sin \phi \hat{e}_x + \cos \phi \hat{e}_y.
\end{eqnarray}
Assume CRF can be obtained by rotating ORF with an angle $\psi$ counterclockwise around $\hat{w}$ axis, the two polarization axes $\hat{p}$ and $\hat{q}$ of the gravitation radiation can be written as
\begin{eqnarray}\label{n7}
&&\hat{p}=\cos \psi\hat{\theta}-\sin\psi\hat{\phi},  \nonumber \\
&&\hat{q} = \sin \psi \hat{\theta}+ \cos \psi \hat{\phi}.
\end{eqnarray}

Based on the above analysis, GW signal propagating along $\hat{k}$ can be equivalently written in ORF and CRF as
\begin{eqnarray}\label{n8}
{\overset{\lower0.5em\hbox{$\smash{\scriptscriptstyle\leftrightarrow}$}} {h} }(t)\!\!\!\!\!\!&&\equiv {h_{\rm{CRF} + }}(t){\varepsilon ^ + } + {h_{\rm{CRF} \times }}(t){\varepsilon ^ \times } \nonumber\\
&&={h_{ + }}(t){e^ + } + {h_{\times }}(t){e^ \times }.
\end{eqnarray}
Here, ${\varepsilon ^ + } \equiv \hat{p} \otimes \hat{p}-\hat{q} \otimes \hat{q}$, ${\varepsilon ^ \times } \equiv\hat{p} \otimes \hat{q}+\hat{q} \otimes \hat{p}$, ${e^ + } \equiv\hat{\theta}  \otimes \hat{\theta}  - \hat{\phi}  \otimes \hat{\phi}$, and ${e^ \times } \equiv \hat{\theta}  \otimes \hat{\phi}  +\hat{\phi}  \otimes \hat{\theta}$ are, respectively, the basis tensors for the two frames, and ${h_{\rm{CRF} + }}(t)$, ${h_{\rm{CRF} \times }}(t)$, ${h_{ + }}(t)$, and ${h_{\times }}(t)$ are the corresponding GW amplitudes. The polarization components in time domain can be written as \cite{1}
\begin{eqnarray}\label{n10}
&&\!\!\!\!\!{h_{\!\text{+}\!}}(t)\!\!=\!\! H({\frac{{1 + \cos^2\iota }}{2}\!\cos 2\psi \cos \Omega t + \cos \iota \sin 2\psi \sin \Omega t} ), \nonumber \\
&&\!\!\!\!\!{h_{\!{\times}\!}}(t)\!\!=\!\!H({\!-\!\frac{{1 + \cos^2\iota }}{2}\!\sin 2\psi\!\cos\Omega t\!+\!\cos\iota \!\cos 2\psi\!\sin \Omega t}),\nonumber \\
\end{eqnarray}
where $\iota$ is the inclination angle of source orbital plane with respect to the $\hat{p}-\hat{q}$ plane. $H=2G^2M_1M_2/R\tilde{D}$ is the GW strain, which can be written as \cite{9}
\begin{eqnarray}\label{n10a}
H\!\!\approx\!\!6.4\!\!\times\!\!10^{-23}\!(\!\frac{M_1}{0.55M_\odot}\!)\!(\!\frac{M_2}{0.27M_\odot}\!)\!(\!\frac{5\rm{kpc}}{\tilde{D}}\!)\!(\!\frac{6.6\!\!\times\!\!10^{4}\rm{km}}{R}\!)
\end{eqnarray}
 for the J0806 wave source in TianQin mission. Here, $M_1$ and $M_2$ are the masses of two stars, $\tilde{D}$ is the distance of the GW source from the sun, and $R$ is the distance between two stars. Making a Fourier transform for Eq. ({\ref{n10}}), one can derive the components in frequency domain as \cite{26}
\begin{eqnarray}\label{n11}
&&{h_ + }(\Omega )\!\!=\!\!H( {\frac{{1 + \cos^2\iota }}{2}\cos 2\psi  - i\cos \iota \sin 2\psi } ), \nonumber \\
&&{h_ \times }(\Omega )\!\!=\!\!H( { - \frac{{1 +\cos^2\iota }}{2}\sin 2\psi  - i\cos \iota \cos 2\psi } ).
\end{eqnarray}

As GW produces spacetime ripples and passes through the space-borne detectors, it can be measured by Doppler shifts of laser frequency. A null space-time element corresponding to a light path can be written as
\begin{equation}\label{n11a}
0=ds^2=c^2dt^2-dx^2-dy^2-dz^2-h_{ij}dx^idx^j
\end{equation}
with the GW perturbation $h_{ij}$. The Doppler shift of laser frequency for one arm of the space-borne detectors has been shown by these works \cite{14,21,26}. Assuming a laser beam from point $A$ is received by point $B$, the beam starts at $t = {t_1}$ from the position $\vec r({t_1}) = {\vec r_A}$, traveling toward $\vec r({t_2}) = {\vec r_B}$, and the light path is $L \equiv c(t_2 - {t_1})$. The position vector of the beam at moment $t$ can be expressed as $\vec r(t) \equiv {\vec r_A} +c (t - {t_1})\hat n$ with $\hat n$ being the unit vector of photon propagation. Therefore, the GW polarization components projected in the beam-propagation direction can be rewritten as \cite{21}
\begin{equation}\label{n12}
h(t)\equiv h_{ij} n^i n^j = {h_ + }(t){\xi _ + } + {h_ \times }(t){\xi _ \times }
\end{equation}
with
\begin{eqnarray}\label{n13}
&&{\xi _ + } = {(\hat{\theta}  \cdot \hat{n})^2} - {(\hat{\phi}  \cdot \hat{n})^2}, \nonumber \\
&&{\xi _ \times } = 2(\hat{\theta}  \cdot \hat{n})(\hat{\phi}  \cdot \hat{n}).
\end{eqnarray}
Two methods can be used to model the measured result due to GW perturbation: one is integrating the line element along photon path between the two points \cite{14,21}, and another is calculating the Doppler shift of the photon propagating from $A$ to $B$ \cite{26}. This one-way signal response for a beam with the fundamental frequency ${\nu _0}$ has been given as \cite{25}
\begin{eqnarray}\label{n14}
\frac{{\delta \nu(\Omega )}}{{{\nu_0}}}\!\!\!\!\!\!&&=\!\!\frac{{ h(\Omega )}}{{2(1\!\!-\!\!\hat{k} \cdot \hat{n})}}{e^{i\Omega \frac{L+\hat{k} \cdot \vec{r}_A}{c}}}{e^{ - i\Omega {t_2}}}\left[ {1 - {e^{ - i\frac{{\Omega L}}{c}(1 - \hat{k} \cdot \hat{n})}}} \right] \nonumber \\
&&\equiv {F_ + }(\Omega ){h_ + }(\Omega ) + {F_ \times }(\Omega ){h_ \times }(\Omega ),
\end{eqnarray}
where ${F_ + }(\Omega )$ and ${F_ \times }(\Omega )$ are the transfer functions of the GW amplitudes in two polarization directions. Now, we will calculate the transfer function for the Michelson combination $X$. The position vectors of the three SCs (see Fig.\,\ref{fig2}) can be rewritten as
\begin{eqnarray}\label{n15}
&&\vec{r}_1 = 0, \nonumber \\
&&\vec{r}_2= L(\cos \frac{\gamma }{2},\sin \frac{\gamma }{2},0), \nonumber \\
&&\vec{r}_3= L(\cos \frac{\gamma }{2}, - \sin \frac{\gamma }{2},0).
\end{eqnarray}

Combining Eqs.\,(\ref{n14}) and (\ref{n15}), one can obtain the transfer functions, e.g., the transfer functions for ${U^1}$ and ${V^1}$ beams can be written as
\begin{eqnarray}\label{n16}
&&\!\!\!\!\!\!\!\!\!\!\!\!{F_{U^1\!,\!+\!,\!\times }}(\Omega )\!\!=\!\!\frac{{{e^{i\Omega (L - \hat{w} \cdot {\vec{r}_3})/c}}}}{{2(1 - \hat{w} \cdot \frac{\vec{r}_3}{L})}}\!\!\left[ {1\!\!-\!\!{e^{ - i\Omega (L - \hat{w} \cdot \vec{r}_3)/c}}} \right]\!\!{\xi _{2;\!+\!,\!\times}}, \nonumber \\
&&\!\!\!\!\!\!\!\!\!\!\!\!{F_{V^1\!,\!+\!,\!\times}}(\Omega )\!\!=\!\!\frac{{\!-{e^{i\Omega (L - \hat{w}\cdot\!{\vec{r}_2})/c}}}}{{2(1\!\!-\!\!\hat{w} \!\cdot\!\frac{\vec{r}_2}{L})}}\!\!\left[ {1\!\!-\!\!{e^{\!- i\Omega(L - \hat{w}\cdot\vec{r}_2)/c}}} \right]\!\!{\xi _{3;\!+\!,\!\times}}\!,
\end{eqnarray}
where the directional functions are
\begin{eqnarray}\label{n17}
\!\!\!\!\!&&\!\!\!\!\!\!\!\!\!\!{\xi _{2; + }} = {\cos ^2}\theta {\cos ^2}\tilde \phi - \sin^2\tilde \phi,\quad {\xi _{2; \times }} =  - \cos \theta \sin2\tilde \phi, \nonumber \\
\!\!\!\!\!&&\!\!\!\!\!\!\!\!\!\!{\xi _{3; + }} = {\cos ^2}\theta {\cos ^2}\underset{\raise0.3em\hbox{$\smash{\scriptscriptstyle\thicksim}$}}{\phi } - \sin^2\underset{\raise0.3em\hbox{$\smash{\scriptscriptstyle\thicksim}$}}{\phi },\quad{\xi _{3; \times }} =  - \cos \theta \sin2\underset{\raise0.3em\hbox{$\smash{\scriptscriptstyle\thicksim}$}}{\phi }
\end{eqnarray}
with
\begin{eqnarray}\label{n24}
\underset{\raise0.3em\hbox{$\smash{\scriptscriptstyle\thicksim}$}}{\phi } \equiv\phi  - \frac{\gamma }{2}\;\;\;,\;\;\;
\tilde \phi  \equiv \phi  + \frac{\gamma }{2}.
\end{eqnarray}
Similarly, the transfer functions of the other beams can be also derived. First, consider the metric of a purely plus-polarized plane gravitational wave, the signal in a single interferometer arm between SC1 and SC2 can be obtained.

Thus, GW signal can be finally expressed as
\begin{equation}\label{n18}
{h}(\Omega ) \equiv F_ +(\Omega ){h_ + }(\Omega ) + F_ \times(\Omega ){ h_ \times }(\Omega )
\end{equation}
for the $X$ data combination of four-link configurations, where the total transfer function is
\begin{eqnarray}\label{n19}
&&F_{+,\times}(\Omega ) = \sum\limits_{i = 1}^3 {{p_i}{F_{V^i; {+,\times}}} + {q_i}{F_{U^i;{+,\times}}}}.
\end{eqnarray}

\begin{figure}[!t]
\includegraphics[width=0.5\textwidth]{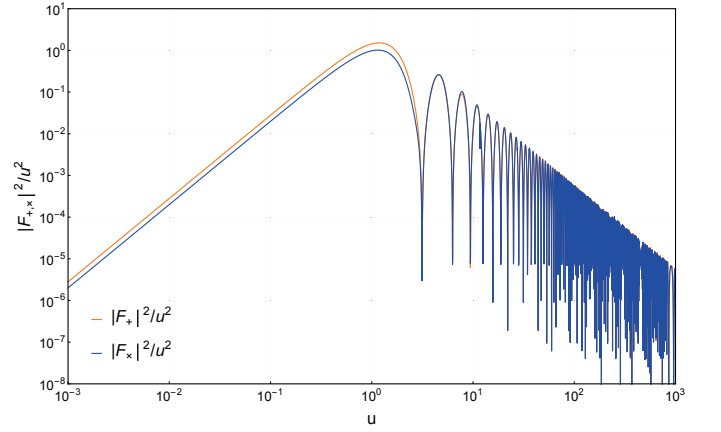}
\caption{ The transfer functions for $+$ and $\times$ polarizations are shown as a function of the variable $u$ for all-sky averaged GW sources, when the angle between the two interference arms is $\lambda=\pi/3$. The orange and blue curves represent the averaged antenna GW response functions of plus and cross polarizations, respectively.}.
\label{fig3}
\end{figure}

Here, we have neglected the small GW signal contribution from the beam $Z^i$. Combining Eqs.\,(\ref{n3}), (\ref{n14}), and (\ref{n19}), we derive the transfer function as
\begin{eqnarray}\label{n20}
\!\!\!\!F_ {+,\times}\!\!=&&\!\!\!\!\!\!\!(1 \!\!-\! {e^{2iu}})\frac{{ \!- {e^{iu(1\! - \sin \!\theta\! \cos \underset{\raise0.3em\hbox{$\smash{\scriptscriptstyle\thicksim}$}}{\phi })}}}}{{2(1 -\!\! \sin \theta \cos \underset{\raise0.3em\hbox{$\smash{\scriptscriptstyle\thicksim}$}}{\phi } )}}\!\!\left[\! {1\! - \! {e^{\! - \!iu(1 \!- \!\sin \theta \cos \underset{\raise0.3em\hbox{$\smash{\scriptscriptstyle\thicksim}$}}{\phi })}}}\! \right]\!\!\xi_{3;+,\!\times}\nonumber \\
&&\!\!\!\!\!\!\!-\! {e^{iu}}\!(1\!\! -\!\! {e^{2iu}})\frac{{\! -\! {e^{iu}}}}{{\!2(1\! + \!\sin \theta \cos \tilde \phi)}}\!\!\left[\! {1 \!-\! {e^{\! -\! iu(1 \!+\! \sin \theta \cos \tilde \phi)}}} \!\right]\!\!\xi_{2;+\!,\times} \nonumber \\
&&\!\!\!\!\!\!\!+(1 \!\!-\! {e^{2iu}})\frac{{\!{e^{iu(1\!- \sin \!\theta\! \cos \tilde \phi)}}}}{{2(1\!\!-\!\!\sin \theta \cos \tilde \phi )}}\!\!\left[\!{1\!-\!{e^{\!-\!iu(1 - \sin \theta \cos\!\tilde \phi)}}}\!\right]\xi_{2;\!+\!,\!\times} \!\!\nonumber \\
&&\!\!\!\!\!\!\! -\!\! {e^{iu}}\!(1\!\! -\!\! {e^{2iu}})\frac{{{e^{iu}}}}{{2(1\!+\!\sin \theta \cos \underset{\raise0.3em\hbox{$\smash{\scriptscriptstyle\thicksim}$}}{\phi })}}\!\left[\!{1\!-\!{e^{ - iu(1 + \sin \theta \cos \underset{\raise0.3em\hbox{$\smash{\scriptscriptstyle\thicksim}$}}{\phi })}}}\!\right]\!\xi_{3;\!+\!,\!\times}.\nonumber \\
\end{eqnarray}
Therefore, the all-sky averaged transfer function is
\begin{eqnarray}\label{n20a}
\mathcal{R}(u)=\frac{1}{8 \pi}\int_{-\pi}^{\pi }d\phi\int_0^{\pi }\sin \theta d\theta ({{{\left|{F_ + }+{F_ \times }\right|}^2}}).
\end{eqnarray}
The transfer functions for $+$ and $\times$ polarizations can be numerical simulated, shown in Fig.\,\ref{fig3}. It is found that the transfer functions are different for different polarizations, and the low-frequency limits are $\frac{14u^4}{5}$ and $\frac{10u^4}{5}$ for $+$ and $\times$ modes, respectively.

\subsection{The equivalence of calculating transfer function with the two methods}\label{subsection3.3}

For the two methods discussed above, the first one calculates the transfer function in the GW frame, which makes the contribution of cross-polarized GW vanished and the contributions of the plus-polarized GW for the two interference arms kept \cite{31,34,35,32,33}; while the case of the $+$ and $\times$ polarization contributions with second method is different. Therefore, it is necessary to study whether the two methods coincide or not.

Based on Sec. {\ref{subsection3.2}}, the function $F_ {+,\times}$ in Eq.\,({\ref{n20}}) can be divided into $F_ {+,\times}^{(1)}$ and $F_ {+,\times}^{(2)}$ for the two interference arms. The transfer functions of plus polarization for arm 1 and arm 2 can be, respectively, expressed as
\begin{align}\label{A4}
F_+^ {(1)} \!\!=&\frac{{-e^{2iu}}(1\!\!-\!\!{e^{2iu}})}{2}[{(1\!\!-\!\!{\mu _1})\!\!+\!\!2{\mu _1}{e^{ \!\!- iu(1 + {\mu _1})}}\!\!-\!\!(1\!\!+\!\!{\mu _1}){e^{ - 2iu}}}],\nonumber\\
F_+^ {(2)} \!\!=&\frac{{-e^{2iu}}(1\!\!-\!\!{e^{2iu}})}{2}[{(1\!\!-\!\!{\mu _2})\!\!+\!\!2{\mu _2}{e^{ \!\!- iu(1 + {\mu _2})}}\!\!-\!\!(1\!\!+\!\!{\mu _2}){e^{ - 2iu}}}]\nonumber\\
&\frac{{\sin ^2}{\theta _2} - 2{(\sin \varepsilon \sin \gamma )^2}}{{\sin ^2}{\theta _2}}.
\end{align}
Similarly, the transfer functions of cross polarization for the two arms can be obtained as
\begin{align}\label{A6}
F_\times^ {(1)} \!\!=&0,\nonumber\\
F_\times^ {(2)} \!\!=&\frac{{-e^{2iu}}(1\!\!-\!\!{e^{2iu}})}{2}\!\![{(1\!\!-\!\!{\mu _2})\!\!+\!\!2{\mu _2}{e^{ \!\!- iu(1 + {\mu _2})}}\!\!-\!\!(1\!\!+\!\!{\mu _2}){e^{ - i2u}}}]\nonumber\\
&\frac{2(\cos {\theta _1}\cos \varepsilon \sin \gamma  - \sin {\theta _1}\cos \gamma )( - \sin \varepsilon \sin \gamma )}{{\sin ^2}{\theta _2}}.
\end{align}
Therefore, the integrated function of Eq.\,({\ref{n20a}}) can be obtained as
\begin{eqnarray}\label{m8}
&&\!\!\!\!\!\!\frac{1}{{2}} {\left| {{F_+}+{F_\times}} \right|^2}\nonumber\\
\!\!\!\!\!\!=&&\!\!\!\!\!\!\frac{1}{2}{\left| {F_+^{(1)} } \right|^2}\!\!+\!\!\frac{1}{2}{\left| {F_2^{(+)} } \right|^2} \!\!+\!\!\frac{1}{2}({F_+^{(1)} F{{_+^{(2)} }^*}\!\!+\!\!F_+^{(2)} F{{_+^{(1)} }^*}})\!\!+\!\!\frac{1}{2}{\left| {F_\times } \right|^2},\nonumber\\
\end{eqnarray}
where the factor $\frac{1}{2}$ comes from the integral of polarization angle $\psi$.

Based on Sec. {\ref{subsection3.1}}, Eqs.\,(\ref{A4}) and (\ref{A6}), the integrand in Eq.\,({\ref{m3}}) can be rewritten as
\begin{eqnarray}\label{m6}
\!\!\!\!\!\!\!\!{\left| {{z_1}\!\!+\!\!{z_2}} \right|^2}\!\!=&&\!\!\!\!\!\!{\left| \cos 2{\psi _1} F_+^ {(1)} \right|^2}\!\!+\!\! {\left| \frac{\cos 2\psi _2\sin ^2 \theta _2 F_+^ {(2)} }{\sin^2 \theta _2\!\!-\!\!2(\sin \varepsilon \sin \gamma )^2} \right|^2}\nonumber\\
&&\!\!\!\!\!\!+\frac{\cos2 \psi _1 \cos 2 \psi _2 \sin ^2 \theta _2}{\sin^2 \theta _2 - 2(\sin \varepsilon \sin \gamma )^2}[F_+^{(1)} F_+^{{(2)}*}\!\!+\!\!F_+^{{(1)}*}F_+^{(2)} ].\nonumber\\
\end{eqnarray}
According to the relationship of the polarization angle $\psi_2=\psi_1+\alpha$ and $\sin \alpha=\sin \gamma \sin \varepsilon /\sin \theta_2$ ($\alpha$ is the angle between the plane containing $\hat{w}$ and arm 1 and the plane containing $\hat{w}$ and arm 2), we average the polarization angle of Eq.\,({\ref{m6}}) and obtain
\begin{eqnarray}\label{m7}
&&\frac{1}{{2\pi }}\int\limits_0^{2\pi }{d\psi } {\left| {{z_1}+{z_2}} \right|^2}\nonumber\\
\!\!\!\!\!\!=&&\!\!\!\!\!\!\frac{1}{2}{\left| {F_+^ {(1)} } \right|^2} + \frac{1}{2}{\left| {F_+^ {(2)}} \right|^2} + \frac{1}{2}({F_+^{(1)} F{{_+^ {(2)} }^*} + F_+^ {(2)} F{{_+^ {(1)} }^*}})\nonumber\\
&&\!\!\!\!\!\!+ \frac{1}{2}{\left| {\frac{4\sin \varepsilon\!\sin \gamma(\cos {\theta _1}\cos \varepsilon \sin \gamma\!\!-\!\!\sin {\theta _1}\cos \gamma ){F_+^ {(2)} }}{{{{\sin }^2}{\theta _2}\!\!-\!\!2{{(\sin \varepsilon \sin \gamma )}^2}}}} \right|^2}.\nonumber\\
\end{eqnarray}

Based on the above discussion, the difference of the transfer functions in these two methods is
\begin{eqnarray}\label{m9}
&&\!\!\!\!\!\!\!{\left| {F_\times } \right|^2}-{\left| {\frac{4\sin \varepsilon\!\sin \gamma(\cos {\theta _1}\cos \varepsilon \sin \gamma\!\!-\!\!\sin {\theta _1}\cos \gamma ){F_+^ {(2)} }}{{{{\sin }^2}{\theta _2}\!\!-\!\!2{{(\sin \varepsilon \sin \gamma )}^2}}}} \right|^2}\nonumber\\
&&\!\!\!\!\!\!\!=0.
\end{eqnarray}
Thus, the two methods for calculating transfer function are proved equivalent. In the following section, we focus on deriving the sensitivity function with the second method and update the analytic expression of the all-sky averaged transfer function.

\section{Sensitivity function}\label{section4}

Based on the above discussions, we can further calculate the sensitivity function, and apply it to the typical space-borne detectors: LISA and TianQin. In general, since we do not know where GWs comes form, we do not know the exact information about the position of the GW source. Assuming a uniform source distribution over sphere, it is reasonable to adopt an inclination-, polarization-, and sky-averaged sensitivity curve \cite{13,15}. Firstly, averaging the polarization and inclination, the PSD of the GW signal emitted by the sources in a certain direction can be obtained as
\begin{eqnarray}\label{n21}
\!\!\!\!\!\!S_h(f)\!\!\!\equiv\!\!\frac{T}{{4\pi }}\!\!\int_0^{\pi } \!\!\!\!\sin\!\iota d\iota\!\!\int_0^{2\pi }\!\!\!\!{{{\left|\!{{{h^{(X)}}(\Omega )}}\!\right|}^2}}\!\! d\psi \!\!=\!\!\frac{2}{5}\!TH^2(\!{{{\left|\!{F_ + ^{(\!X\!)}}\!\right|}^2}\!\!+\!\!{{\left|\!{F_ \times ^{(\!X\!)}}\!\right|}^2}}\!\!\!)
\end{eqnarray}
with $T$ being the observation time. Further making an all-sky average to achieve the GW signal of the all-sky sources is not easy, hence a lot of related works just give the mathematic simulation. Here, we appropriately adopt a series of coordinate transformations transferring the spherical surface integral to a plane one, which simplifies the calculation procedure. Finally, the PSD of the GW signal for the all-sky sources can be obtained as
\begin{eqnarray}\label{n21a}
\!\!\!\!\!\!\!\!\tilde{S}_h(f)\!\!\equiv\!\!\frac{1}{{4\pi }}\!\!\int_{-\pi}^{\pi }\!\!\!d\phi\!\!\int_0^{\pi }\!\!\!\!S_h(f)\!\sin \theta d\theta\!\!=\!\!\frac{4}{5}\!TH^2\!\mathcal{R}(u).
\end{eqnarray}

Therefore, the signal-to-noise ratio (SNR) for the Michelson data combination $X$ at a given GW frequency $f$ can be defined as
\begin{equation}\label{n22}
{\rm{SNR}} \equiv \sqrt {\frac{{{\tilde{S}_h}(f)}}{{S_N^{\!(\!X\!)\!}}(f)}},
\end{equation}
and the sensitivity function
\begin{eqnarray}\label{n23}
\!\!\!\!\!\!S(f)\!\!&&\!\!\!\!\!\!\equiv \!\!5\sqrt {\frac{{S_N^{\!(\!X\!)\!}}(f)}{{{\tilde{S}_h}(f)/H^2}}}\nonumber\\
&&\!\!\!\!\!\!=\!\!5\frac{{\sqrt {{\!\!(8\sin^2 2u+32\sin^2 u) \frac{{{L^2}s_a^2}}{{{u^2}{c^4}}}\!\!+\!\!16\frac{{{u^2}s_x^2}}{{{L^2}}}}} }}{{\sqrt {\frac{4T}{5}\mathcal{R}(u)}}},
\end{eqnarray}
where 5 represents the SNR in 1-year observation time and $\mathcal{R}(u)\equiv \frac{1}{2}\sin^2 u [f(u) + g(u)]$ is calculated in the Appendix. Combining the above analysis, we will calculate the sensitivity functions of the equilateral triangle-formation space-borne GW detectors, LISA and TianQin. As for $\gamma=\pi /3$, the below simplification can be derived as
\begin{eqnarray}\label{n36}
\mathcal{R}(u)\!\!=&&\!\!\!\!\!\!\!\!2\{\frac{{5\!+\!12\ln2}}{3}\!\!-\!\!\frac{4}{{{u^2}}}-\!4[{\rm{Ci}}(2u)\!\!-\!\!{\rm{Ci}}(u)]\}\sin^2 u\!\!\nonumber \\
&&\!\!\!\!\!\!\!\!+\!2(6\ln 3\!-\!\!12\ln 2\!\!+\!\!\frac{1}{3}\!\!-\!\!\frac{4}{{{u^2}}})\!\!\cos 2u \sin^2 u\!\!+\!\!\frac{{10}}{{{u^2}}}\cos u \sin^2u\nonumber \\
&&\!\!\!\!\!\!\!\!+8(\frac{1}{u}\!+\!\frac{2}{{{u^3}}})\cos u \sin^3 u\!-\!2(\frac{7}{u}\!+\!\frac{5}{{{u^3}}})\sin^3 u\!\nonumber \\
&&\!\!\!\!\!\!\!\!-\!12\{\cos(2u)[{\rm{Ci}}(3u)\!+\!{\rm{Ci}}(u)\!-\!2{\rm{Ci}}(2u)]\!\nonumber \\
&&\!\!\!\!\!\!\!\!+\!\sin(2u)[{\rm{Si}}(3u)\!+\!{\rm{Si}}(u)\!-\!2{\rm{Si}}(2u)]\}\sin^2 u.
\end{eqnarray}
This detailed calculation has been presented in the Appendix. Compared with the approximately analytic calculation of the GW transfer function in Refs. \cite{12,30a}, Eq.\,({\ref{n36}}) is more complete and simpler. This transfer function can be approximate to $\frac{12}{5}u^4+\frac{337}{210}u^6$ at the low-frequency limit.

\begin{figure}[!t]
\includegraphics[width=0.5\textwidth]{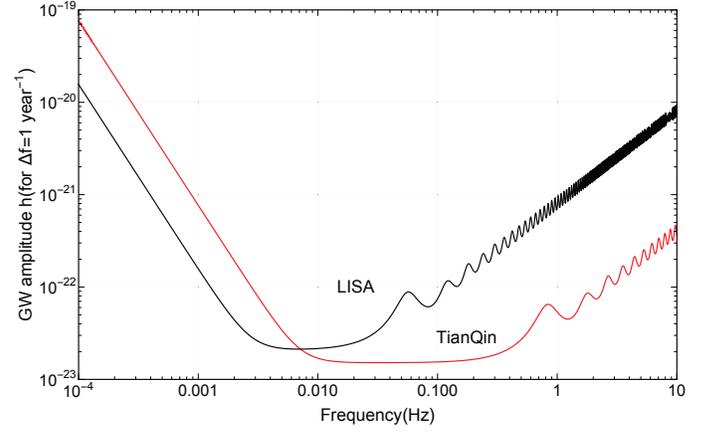}
\caption{ Sensitivity curves for the GW detectors (LISA and TianQin) averaging all-sky sources in the four-link configurations.}
\label{fig4}
\end{figure}

For LISA mission, the arm length is $L=2.5\times10^6$km, and the preliminary goal ASD of the proof mass noise and shot noise are, respectively, $s_a^{\rm{LISA}} = 3 \times {10^{ - 15}}\rm{m{s^{ - 2}}/\sqrt {Hz}}$, $s_x^{\rm{LISA}} = 20 \times {10^{ - 12}}\rm{m/\sqrt {Hz}}$, while the case for TianQin is $L=1.7\times10^5$km, $s_a^{\rm{TQ}} = 1\times{10^{ - 15}}\rm {m{s^{ - 2}}/\sqrt {Hz}}$, and $s_x^{\rm{TQ}} = 1 \times {10^{ - 12}} \rm {m/\sqrt {Hz}}$. Therefore, the sensitivity curves of these two missions can be figured out as Fig. \ref{fig4}, both of which reflect the GW signal emitted by all-sky sources. This figure shows LISA mission is sensitive to the signals at a lower frequency band, while TianQin mission is sensitive to the signals at a higher frequency band.

\begin{figure}[!t]
\includegraphics[width=0.5\textwidth]{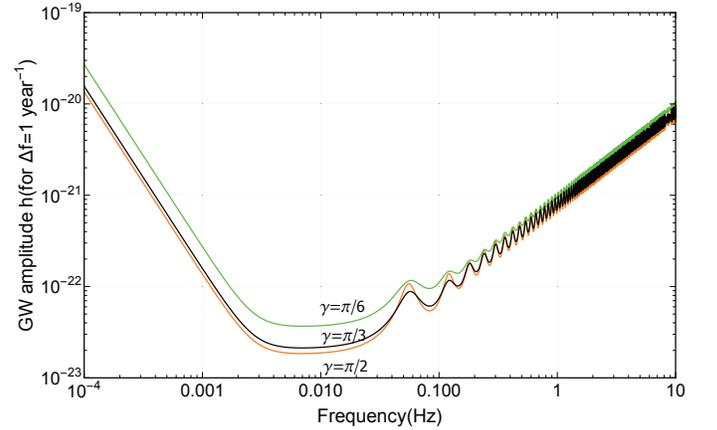}
\caption{ Dependence of the sensitivity curve of LISA mission for Michelson data combination on the angle between two effective interfering arms ($\gamma=\pi/2$, $\pi/3$ and $\pi/6$). }
\label{fig5}
\end{figure}

Essentially, the Michelson data combination only involves two arms, since $p_2=q_3=0$. Similar to ground-based interferometers, one may wonder what the difference of the sensitivity curve will be if the two arms are perpendicular to each other. Or what the case is, if the angle between the two effective interference arms is a smaller one, such as $\pi/6$. Taking LISA mission, e.g., we have made a comparison and found the smaller the angle is, the worse the sensitivity will be (shown by Fig. \ref{fig5}). Moreover, the magnitude of the sensitivity function is inversely proportional to the sinusoidal value of the angle approximately at low frequencies ($u<\pi$), and the sensitivity-curve oscillating amplitude at the high frequencies band ($u>\pi$) decreases with decreasing the value of $\gamma$. This means the formation with $\gamma=\pi/2$ is the best one for the space-borne GW detector with the Michelson data combination-type TDI technique.

\begin{figure}[!t]
\includegraphics[width=0.5\textwidth]{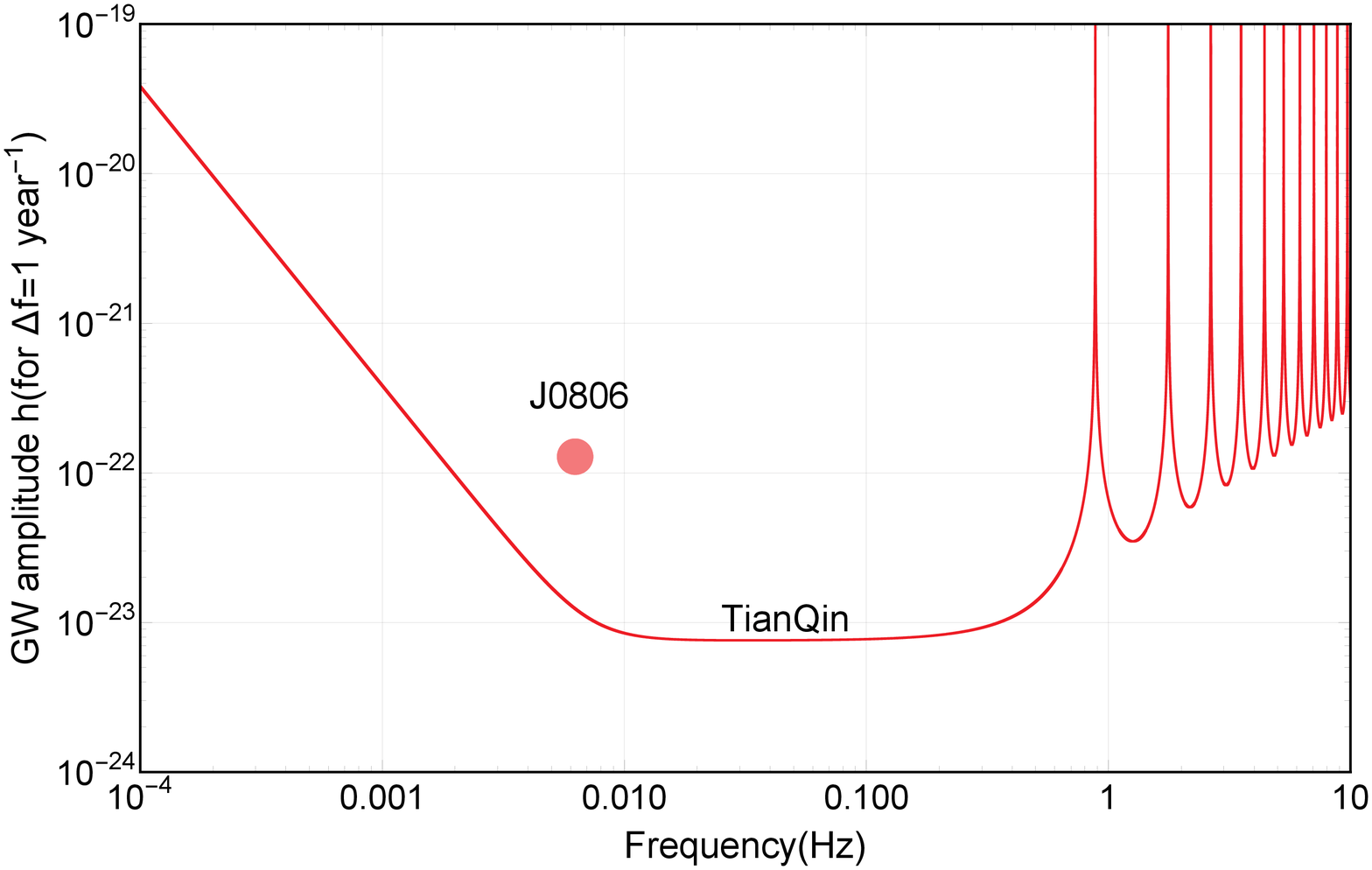}
\caption{ Sensitivity curve of TianQin mission, which focuses on the GW signal emitted by the special source (J0806). The red line and the pink point represent the mission's sensitivity curve and the GW signal, respectively.}
\label{fig6}
\end{figure}

Although TianQin mission similar to LISA mission can detect the GW signal emitted by all-sky sources, its primary goal focuses on the special source J0806, the calculation of whose sensitivity function does not need to perform the step shown in Eq. (\ref{n21a}). In this case, TianQin detector is designed as the plane composed of the three SCs being perpendicular to the GW propagation direction, that is, $\theta=\phi=0$. Thus, according to Eq. (\ref{n25}), the relationship can be obtained as follows:
\begin{eqnarray}\label{n26}
{\left| {\frac{{F_ + ^{(X)}}}{{\sin u}}} \right|^2} + {\left| {\frac{{F_ \times ^{(X)}}}{{\sin u}}} \right|^2} = 16{\sin ^2}u{\sin ^2}\gamma.
\end{eqnarray}
Furthermore, one can get the sensitivity function as
\begin{eqnarray}\label{n27}
\!\!\!\!\!\!{S_{\rm{TQ}}}\!(\!f\!)\!\!&&\!\!\!\!\!\!\equiv \!\!5\sqrt {\!\!\frac{{S_N^{\!(X)\!}}(f)}{{{S_h}(f)/H^2}}}\!\!=\!\!5\frac{{\sqrt {{\!\!32\left(\!{1\!\!+\!\!{\cos^2 u}} \right)\!\frac{{{L^2}s_a^2}}{{{u^2}{c^4}}}\!\!+\!\!16\frac{{{u^2}s_x^2}}{{{L^2}}}}} }}{{\sqrt {\frac{2T}{5}}  \times 4\sin u\sin \gamma }}.
\end{eqnarray}
The corresponding sensitivity curve in Fig. \ref{fig6} shows TianQin is sensitive to GW signal at about 0.001$\sim$0.5Hz and insensitive in the high frequency band, and there are some nodes in some particular frequencies satisfying $u=k\pi$ with $k$ being integer. The signal at these frequencies cannot be detected, which arises from the Michelson data combination that not only cancels the laser phase noise, but also cancels GW signal greatly. Since this phenomenon only occurs at a certain azimuth angle, this node effect may be heavily suppressed for the all-sky source case averaging the sensitivity function over all-around azimuth angles, which agrees with the curve at high frequencies in Fig. \ref{fig4}.

\section{summary}\label{section5}

This paper calculated the transfer function of space-borne GW detectors with the method dividing the GW signal into two polarizations, which is proved to be equivalent with the conventional method. Based on this, we successfully gave a full analytical expression of the sensitivity function with the Michelson data combination-type TDI technique and further applied it in LISA and TianQin missions. Then, we discussed the dependence of the sensitivity function on the angle between both arms of the interference, and found the amplitude of the sensitivity only at the low frequencies ($u<\pi$) is inversely proportional to the sinusoidal value of the angle approximately, and at the high frequencies ($u>\pi$) oscillates more heavily with increasing this angle. This analytic work may guide the experiments to propose that the technique demands and designs a better GW detector.

\section*{ACKNOWLEDGMENTS}
This work is supported by the National Natural Science Foundation of China(Grants No. 91636221 and No. 11805074) and the Postdoctoral Science Foundation of China (Grants No. 2017M620308 and No. 2018T110750).

\section*{APPENDIX: ANALYTIC CALCULATION RELATED TO TRANSFER FUNCTION OF THE GRAVITATIONAL WAVE SIGNAL}

This appendix provides a detailed calculation related to the transfer function of the GW signal with the Michelson data combination. According to Eq. (\ref{n20}), we can derive
\begin{widetext}
\begin{equation}\label{n25}
{\left|\!{\frac{{F_ + ^{(X)}}}{{\sin u}}}\!\right|^2}\!\!\!\!+\!\!{\left|\!{\frac{{F_ \times ^{(X)}}}{{\sin u}}} \!\right|^2}\!\!\!\!=\!{m_1}\!(\!u,\!\theta,\!\phi,\!\gamma ) \!\!+\!\!{m_2}\!(\!u,\!\theta,\!\phi,\!\gamma )\!\!+\!\!{m_3}\!(\!u,\!\theta,\!\phi,\!\gamma ),\tag{A1}
\end{equation}
with
\begin{align}
{m_1}(u,\theta ,\phi ,\gamma )\!\!=\!\!&-\!2\{{1\!+\!\cos (2u)\!+\!2\cos[2u\sin \theta \sin \phi \sin (\gamma /2)]} \}{\sin ^2}\theta (\cos \gamma\!+\!\cos 2\phi )+ 2(3 + \cos 2u)( {{{\sin }^2}\theta {{\cos }^2}\underset{\raise0.3em\hbox{$\smash{\scriptscriptstyle\thicksim}$}}{\phi }\!+\!{{\sin }^2}\theta {{\cos }^2}\tilde \phi } )\nonumber
\end{align}
\begin{align}
{m_2}(u,\theta ,\phi ,\gamma )\!\!=\!\!&+ 2\left( \begin{gathered}
   - \cos [u(1 - \sin \theta \cos \underset{\raise0.3em\hbox{$\smash{\scriptscriptstyle\thicksim}$}}{\phi } )]{(1 + \sin \theta \cos \underset{\raise0.3em\hbox{$\smash{\scriptscriptstyle\thicksim}$}}{\phi } )^2} - \cos [u(1 + \sin \theta \cos \underset{\raise0.3em\hbox{$\smash{\scriptscriptstyle\thicksim}$}}{\phi } )]{(1 - \sin \theta \cos \underset{\raise0.3em\hbox{$\smash{\scriptscriptstyle\thicksim}$}}{\phi } )^2} \hfill \\
   + \left[ {\cos [u(1 - \sin \theta \cos \underset{\raise0.3em\hbox{$\smash{\scriptscriptstyle\thicksim}$}}{\phi } )] + \cos [u(1 + \sin \theta \cos \underset{\raise0.3em\hbox{$\smash{\scriptscriptstyle\thicksim}$}}{\phi } )]} \right](1 - {\sin ^2}\theta {\cos ^2}\underset{\raise0.3em\hbox{$\smash{\scriptscriptstyle\thicksim}$}}{\phi } ) \hfill \\
\end{gathered}  \right) \nonumber\\
&+ 2\left( \begin{gathered}
   - \cos [u(1 - \sin \theta \cos \tilde \phi )]{(1 + \sin \theta \cos \tilde \phi )^2} - \cos [u(1 + \sin \theta \cos \tilde \phi )]{(1 - \sin \theta \cos \tilde \phi )^2} \nonumber \\
   \!\!\!\!\!\!\!\!\!\!\!\!\!\!\!\!\!\!\!\!\!\!\!\!\!\!\!\!\!\!\!\!\!\!\!\!+ \left[ {\cos [u(1 - \sin \theta \cos \tilde \phi )] + \cos [u(1 + \sin \theta \cos \tilde \phi )]} \right](1 - {\sin ^2}\theta {\cos ^2}\tilde \phi ) \nonumber \\
\end{gathered}  \right) \nonumber \\
&- 2\left( {\cos [u(1 - \sin \theta \cos \tilde \phi )] + \cos [u(1 + \sin \theta \cos \underset{\raise0.3em\hbox{$\smash{\scriptscriptstyle\thicksim}$}}{\phi } )]} \right)(1 + \sin \theta \cos \underset{\raise0.3em\hbox{$\smash{\scriptscriptstyle\thicksim}$}}{\phi } )(1 - \sin \theta \cos \tilde \phi ) \nonumber \\
&- 2\left( {\cos [u(1 + \sin \theta \cos \tilde \phi )] + \cos [u(1 - \sin \theta \cos \underset{\raise0.3em\hbox{$\smash{\scriptscriptstyle\thicksim}$}}{\phi } )]} \right)(1 - \sin \theta \cos \underset{\raise0.3em\hbox{$\smash{\scriptscriptstyle\thicksim}$}}{\phi } )(1 + \sin \theta \cos \tilde \phi ) \nonumber\\
&+ 2\left( {\cos [u(1 + \sin \theta \cos \underset{\raise0.3em\hbox{$\smash{\scriptscriptstyle\thicksim}$}}{\phi } )] + \cos [u(1 + \sin \theta \cos \tilde \phi )]} \right)(1 - \sin \theta \cos \underset{\raise0.3em\hbox{$\smash{\scriptscriptstyle\thicksim}$}}{\phi } )(1 - \sin \theta \cos \tilde \phi ) \nonumber \\
&+ 2\left( {\cos [u(1 - \sin \theta \cos \underset{\raise0.3em\hbox{$\smash{\scriptscriptstyle\thicksim}$}}{\phi } )] + \cos [u(1 - \sin \theta \cos \tilde \phi )]} \right)(1 + \sin \theta \cos \underset{\raise0.3em\hbox{$\smash{\scriptscriptstyle\thicksim}$}}{\phi } )(1 + \sin \theta \cos \tilde \phi ) \nonumber\\
{m_3}(u,\theta ,\phi ,\gamma )\!\!=\!\!& - 4\left( \begin{gathered}
\frac{{\cos [u(1 - \sin \theta \cos \tilde \phi )]}}{{1 - \sin \theta \cos \underset{\raise0.3em\hbox{$\smash{\scriptscriptstyle\thicksim}$}}{\phi } }}\frac{{2\sin \theta \cos \tilde \phi }}{{1 - {{\sin }^2}\theta {{\cos }^2}\tilde \phi }} + \frac{{\cos [u(1 + \sin \theta \cos \underset{\raise0.3em\hbox{$\smash{\scriptscriptstyle\thicksim}$}}{\phi } )]}}{{1 + \sin \theta \cos \tilde \phi }}\frac{{ - 2\sin \theta \cos \underset{\raise0.3em\hbox{$\smash{\scriptscriptstyle\thicksim}$}}{\phi } }}{{1 - {{\sin }^2}\theta {{\cos }^2}\underset{\raise0.3em\hbox{$\smash{\scriptscriptstyle\thicksim}$}}{\phi } }} \hfill \\
+\frac{{\cos [u(1 + \sin \theta \cos \tilde \phi )]}}{{1 + \sin \theta \cos \underset{\raise0.3em\hbox{$\smash{\scriptscriptstyle\thicksim}$}}{\phi } }}\frac{{ - 2\sin \theta \cos \tilde \phi }}{{1 - {{\sin }^2}\theta {{\cos }^2}\tilde \phi }} + \frac{{\cos [u(1 - \sin \theta \cos \underset{\raise0.3em\hbox{$\smash{\scriptscriptstyle\thicksim}$}}{\phi } )]}}{{1 - \sin \theta \cos \tilde \phi }}\frac{{2\sin \theta \cos \underset{\raise0.3em\hbox{$\smash{\scriptscriptstyle\thicksim}$}}{\phi } }}{{1 - {{\sin }^2}\theta {{\cos }^2}\underset{\raise0.3em\hbox{$\smash{\scriptscriptstyle\thicksim}$}}{\phi } }} \hfill \\
+\frac{{\cos 2u + \cos[u\sin \theta (\cos \underset{\raise0.3em\hbox{$\smash{\scriptscriptstyle\thicksim}$}}{\phi }  - \cos \tilde \phi )]}}{{(1 - {{\sin }^2}\theta {{\cos }^2}\underset{\raise0.3em\hbox{$\smash{\scriptscriptstyle\thicksim}$}}{\phi })(1 - {{\sin }^2}\theta {{\cos }^2}\tilde \phi )}}2(1 - {\sin ^2}\theta \cos \underset{\raise0.3em\hbox{$\smash{\scriptscriptstyle\thicksim}$}}{\phi } \cos \tilde \phi ) \hfill \\
\!\!\!\!\!\!\!\!\!\!\!\!\!\!\!\!\!\!\!\!\!\!\!\!\!\!\!\!\!\!\!\!\!\!\!\!\!\!\!\!\!\!\!\!\!\!\!\!\!\!\!\!\!\!\!- \frac{{1 + \cos[u\sin \theta (\cos \underset{\raise0.3em\hbox{$\smash{\scriptscriptstyle\thicksim}$}}{\phi }  - \cos \tilde \phi )]}}{{(1 - {{\sin }^2}\theta {{\cos }^2}\underset{\raise0.3em\hbox{$\smash{\scriptscriptstyle\thicksim}$}}{\phi } )(1 - {{\sin }^2}\theta {{\cos }^2}\tilde \phi )}}2(1 + {\sin ^2}\theta \cos \underset{\raise0.3em\hbox{$\smash{\scriptscriptstyle\thicksim}$}}{\phi } \cos \tilde \phi ) \nonumber
\end{gathered}  \right){\cos ^2}\theta {\sin ^2}\gamma \nonumber \\ \tag{A2}
\end{align}
Further, integrating the angle $(\theta,\phi)$ for the whole space, one can obtain
\begin{equation}\label{n29}
\!\!\frac{1}{{4\pi }}\int_0^\pi\!\!{\sin \theta } d\theta{\int_{ - \pi }^\pi } d\phi \left[\!{{{\left| \!{\frac{{F_ + ^{(X)}}}{{\sin u}}} \!\right|}^2} + {{\left| {\frac{{F_ \times ^{(X)}}}{{\sin u}}} \right|}^2}} \right]=f(u)\!+\!g(u). \tag{A3}
\end{equation}
Here, the functions are
\begin{align}\label{n30}
&f(u) = {f_0}(u) + {b_s}{\sin ^2}u + {b_c}{\cos ^2}u, \nonumber \\
&g(u) = {g_0}(u) + {g_s}(u)\sin u + {g_c}(u)\cos u, \tag{A4}
\end{align}
with
\begin{align}\label{n31}
&{f_0}(u) =\frac{4}{3}[(3 + \cos 2u) - \left( {1 + \cos 2u} \right)\cos \gamma ] + 16\cos u(1 - \cos \gamma )\frac{{(2 - {u^2})\sin u - 2u\cos u}}{{{u^3}}} \nonumber \\
&\;\;\;\;\;\;\;\;\;\;\;\;- 4\frac{{\left[ {(2u\sin \frac{\gamma }{2})\cos (2u\sin \frac{\gamma }{2}) - \sin (2u\sin \frac{\gamma }{2})} \right](\cos\gamma  - 3) + {{(2u\sin \frac{\gamma }{2})}^2}\sin (2u\sin \frac{\gamma }{2})(\cos\gamma  - 1)}}{{{{(2u\sin \frac{\gamma }{2})}^3}}},\nonumber \\
&{b_s} = \int\limits_0^{\pi /2} {\frac{{\sin \theta }}{\pi }d\theta } \int\limits_0^\pi  {d\phi } \frac{{16{{\cos }^2}\theta {{\sin }^2}\gamma }}{{(1 - {{\sin }^2}\theta {{\cos }^2}\underset{\raise0.3em\hbox{$\smash{\scriptscriptstyle\thicksim}$}}{\phi } )(1 - {{\sin }^2}\theta {{\cos }^2}\tilde \phi )}},\nonumber
\end{align}
\begin{align}
&{b_c} =\int\limits_0^{\pi /2} {\frac{{\sin \theta }}{\pi }d\theta } \int\limits_0^\pi  {d\phi } \frac{{16{{\cos }^2}\theta {{\sin }^2}\gamma {{\sin }^2}\theta \cos \underset{\raise0.3em\hbox{$\smash{\scriptscriptstyle\thicksim}$}}{\phi } \cos \tilde \phi }}{{(1 - {{\sin }^2}\theta {{\cos }^2}\underset{\raise0.3em\hbox{$\smash{\scriptscriptstyle\thicksim}$}}{\phi } )(1 - {{\sin }^2}\theta {{\cos }^2}\tilde \phi )}} \nonumber \\
&{g_0}(u) =\int\limits_0^{\pi /2} {\frac{{\sin \theta }}{\pi }d\theta } \int\limits_0^\pi  {d\phi } \frac{{16{{\cos }^2}\theta {{\sin }^2}\gamma {{\sin }^2}\theta \cos \underset{\raise0.3em\hbox{$\smash{\scriptscriptstyle\thicksim}$}}{\phi } \cos \tilde \phi }}{{(1 - {{\sin }^2}\theta {{\cos }^2}\underset{\raise0.3em\hbox{$\smash{\scriptscriptstyle\thicksim}$}}{\phi } )(1 - {{\sin }^2}\theta {{\cos }^2}\tilde \phi )}}\cos[u\sin \theta (\cos \underset{\raise0.3em\hbox{$\smash{\scriptscriptstyle\thicksim}$}}{\phi }  - \cos \tilde \phi )]
\nonumber\\
&{g_s}(u) = \int\limits_0^{\pi /2} {\frac{{\sin \theta }}{\pi }d\theta } \int\limits_0^\pi  {d\phi } \frac{{ - 16{{\cos }^2}\theta {{\sin }^2}\gamma \sin \theta }}{{(1 - {{\sin }^2}\theta {{\cos }^2}\underset{\raise0.3em\hbox{$\smash{\scriptscriptstyle\thicksim}$}}{\phi } )(1 - {{\sin }^2}\theta {{\cos }^2}\tilde \phi )}}[\sin (u\sin \theta \cos \underset{\raise0.3em\hbox{$\smash{\scriptscriptstyle\thicksim}$}}{\phi } )\cos \underset{\raise0.3em\hbox{$\smash{\scriptscriptstyle\thicksim}$}}{\phi }  + \sin (u\sin \theta \cos \tilde \phi )\cos \tilde \phi ] \nonumber\\
&{g_c}(u) =\int\limits_0^{\pi /2} {\frac{{\sin \theta }}{\pi }d\theta } \int\limits_0^\pi  {d\phi } \frac{{ - 16{{\cos }^2}\theta {{\sin }^2}\gamma {{\sin }^2}\theta \cos \underset{\raise0.3em\hbox{$\smash{\scriptscriptstyle\thicksim}$}}{\phi } \cos \tilde \phi }}{{(1 - {{\sin }^2}\theta {{\cos }^2}\underset{\raise0.3em\hbox{$\smash{\scriptscriptstyle\thicksim}$}}{\phi } )(1 - {{\sin }^2}\theta {{\cos }^2}\tilde \phi )}}[\cos (u\sin \theta \cos \underset{\raise0.3em\hbox{$\smash{\scriptscriptstyle\thicksim}$}}{\phi } ) + \cos (u\sin \theta \cos \tilde \phi )].\tag{A5}
\end{align}
\end{widetext}
Next, we choose appropriate  reference frames to simplify the calculation of the analytic expressions for the integrals of the above equations. First, by the variable substitution
\begin{align}\label{n33}
\left\{ \begin{gathered}
x = \sin \theta \cos \phi  \hfill \\
y = \sin \theta \sin \phi \hfill \\
\end{gathered}  \right.,\tag{A6}
\end{align}
spherical integral of a unit-radius sphere can be equivalent as a circular surface integral. In this case, $\sin \theta d\theta d\phi=dxdy/\cos\theta$, and the integral region of a unit hemispherical surface [$\theta \in (0,\pi/2), \phi \in (0,2\pi) $] is changed as ${x^2} + {y^2} \leqslant 1$. Through a further transformation
\begin{align}\label{n33c}
\left\{ \begin{gathered}
\tilde x \equiv \sin \theta \cos \underset{\raise0.3em\hbox{$\smash{\scriptscriptstyle\thicksim}$}}{\phi }  = x\cos \frac{\gamma }{2} + y\sin \frac{\gamma }{2} \hfill \\
\tilde y \equiv \sin \theta \cos \tilde \phi  = x\cos \frac{\gamma }{2} - y\sin \frac{\gamma }{2}\hfill \\
\end{gathered}  \right.,\tag{A7}
\end{align}
the circular surface integral is stretched as an elliptic integral, $dxdy=d\tilde x d\tilde y/\sin\gamma$, and the integral region of a circular surface is changed as $\frac{{\tilde x}^2}{2\cos^2 \gamma/2}+\frac{{\tilde y}^2}{2\sin^2 \gamma/2}\leqslant 1$. Thus, the calculation of Eq. (\ref{n31}) can be simplified. For example, $b_c$ can be written as
\begin{widetext}
\begin{align}\label{n33a}
{b_c} = \frac{8}{\pi }\int\limits_{ - 1}^1\!\!{d\tilde y} \frac{{\tilde y}}{{1 - {{\tilde y}^2}}}\int\limits_{\tilde y\cos \gamma  - \sin \gamma \sqrt {1 - {{\tilde y}^2}} }^{\tilde y\cos \gamma  + \sin \gamma \sqrt {1 - {{\tilde y}^2}} }{d\tilde x} \frac{{\sqrt {{{\sin }^2}\gamma  - {{\tilde x}^2} - {{\tilde y}^2} + 2\tilde x\tilde y\cos \gamma } \tilde x}}{{1 - {{\tilde x}^2}}},\tag{A8}
\end{align}
where the second integral can be calculated as
\begin{align}\label{n33b}
&\int\limits_{\tilde y\cos \gamma  - \sin \gamma \sqrt {1 - {{\tilde y}^2}} }^{\tilde y\cos \gamma  + \sin \gamma \sqrt {1 - {{\tilde y}^2}} }{d\tilde x} \frac{{\sqrt {{{\sin }^2}\gamma  - {{\tilde x}^2} - {{\tilde y}^2} + 2\tilde x\tilde y\cos \gamma } \tilde x}}{{1 - {{\tilde x}^2}}}=\pi \left[ {\frac{{\left| {\tilde y + \cos \gamma } \right| - \left| {\tilde y - \cos \gamma } \right|}}{2} - \tilde y\cos \gamma } \right].\nonumber \\ \tag{A9}
\end{align}
Therefore, Eq. ({\ref{n33a}}) can be given as
\begin{align}\label{n33d}
\!\!\!\!{b_c} =\!\!8(1\!-\!\cos \gamma ){\rm{ln}}\frac{{1\!+\!\cos \gamma }}{{1\!-\! \cos \gamma }}\!\!-\!\!16\cos \gamma \ln \frac{2}{{1\!+\!\cos \gamma }}.\tag{A10}
\end{align}
Similarly, the others functions can be derived as
\begin{align}
&{b_s}\!=\! 8(1 - \cos \gamma ){\rm{ln}}\frac{{1 + \cos \gamma }}{{1 - \cos \gamma }} + 16\ln \frac{2}{{1 + \cos \gamma }},\nonumber \\
&{g_0}(u)\!=\!32\frac{{\sin (2u{{\sin }^2}\frac{\gamma }{2})\!\!-\!\!\sin \frac{\gamma }{2}\sin (2u\sin \frac{\gamma }{2})}}{u}-\!\!16{\cos ^2}\frac{\gamma }{2}\{ \cos(2u)\![ {{\text{Ci}}(2u\!\!+\!\!2\sin\!\frac{\gamma }{2}u)\!\!+\!\!{\text{Ci}}(2u\!\!-\!\!2\sin\!\frac{\gamma }{2}u)\!\!-\!\!2{\text{Ci}}(2u\!\!-\!\!2{{\sin }^2}\!\frac{\gamma }{2}u)} ]\nonumber \\
&\;\;\;\;\;\;\;\;\;\;\;\;+\!\!\sin (2u)\![ {{\text{Si}}(2u\!\!+\!\!2\sin\!\frac{\gamma }{2}u)\!\!+\!\!{\text{Si}}(2u\!\!-\!\!2\sin \!\frac{\gamma }{2}u)\!\!-\!\!2{\text{Si}}\!(2u\!\!-\!\!2{{\sin }^2}\!\!\frac{\gamma }{2}u)} ]\}\!\!+\!\!32{\sin ^2}\!\frac{\gamma }{2}\left[ {{\text{Ci}}(2\sin \frac{\gamma }{2}u)\!\!-\!\!{\text{Ci}}(2{{\sin }^2}\frac{\gamma }{2}u)} \right], \nonumber\\
&{g_s}(u) =\!-\!32\frac{{\cos (u\cos \gamma )\!-\!\cos u}}{u}\!-\!16\cos \gamma \{{\cos u\left[ {{\text{Si(}}u\!+\!u\cos \gamma {\text{)}}\!-\!{\text{Si(}}u\!-\!u\cos \gamma {\text{)}}} \right]\!+\!\sin u\left[ {\!-\!{\text{Ci(}}u\!+\!u\cos \gamma {\text{)}}\!+\!{\text{Ci(}}u\!-\!u\cos \gamma {\text{)}}} \right]} \}\nonumber \\
&\;\;\;\;\;\;\;\;\;\;\;\;+\!16\{{\cos u\left[ {{\text{2Si(2}}u{\text{)}}\!-\!{\text{Si(}}u\!+\!u\cos \gamma {\text{)}}\!-\!{\text{Si(}}u\!-\!u\cos \gamma {\text{)}}} \right]\!-\!\sin u\left[ {2{\text{Ci(2}}u{\text{)}}\!-\!{\text{Ci(}}u\!+\!u\cos \gamma {\text{)}}\!-\!{\text{Ci(}}u\!-\!u\cos \gamma {\text{)}}} \right]} \}, \nonumber \\
&{g_c}(u)\!\!= 32\!\cos \gamma \!\frac{{\sin (u\!\cos \gamma )\!-\!\cos \gamma \sin u}}{u}\!-\!16\{ {\cos u\left[ {{\text{Ci(}}u\!\!+\!\!u\cos \gamma {\text{)}}\!-\!{\text{Ci(}}u\!-\!u\cos \gamma {\text{)}}} \right]\!+\!\sin u\left[ {{\text{Si(}}u\!+\!u\cos \gamma {\text{)}} \!-\! {\text{Si(}}u\!-\!u\cos \gamma {\text{)}}} \right]} \}, \nonumber \\
&\;\;\;\;\;\;\;\;\;+\!16\cos \!\gamma \{ {\cos u\left[ {2{\text{Ci(2}}u{\text{)}} \!\!-\!\!{\text{Ci(}}u \!+\! u\cos \gamma {\text{)}} \!\!-\!\!{\text{Ci(}}u \!-\! u\cos \gamma {\text{)}}} \right] \!\!+\!\! \sin u\left[ {{\text{2Si(2}}u{\text{)}}\!\!-\!\!{\text{Si(}}u \!+\! u\cos \gamma {\text{)}}\!\!-\!\!{\text{Si(}}u \!\!-\!\! u\cos \gamma {\text{)}}} \right]} \},\tag{A11}
\end{align}
thus we can obtain the analytic expression of Eq. ({\ref{n30}}) for arbitrary value of the angle $\gamma$. Then, we simplify the above functions and get the following results:
\begin{align}\label{n35}
\!\!\!\!\!\!\!\!\!\!\!\!\!\!\!\!\!\!\!\!&f(u)\!\!=\!\!\frac{4}{3}[(3\!+\!\cos 2u)\!-\!\left( {1\!+\!\cos 2u} \right)\cos \gamma ]\!+\!16\cos u(1\!-\!\cos \gamma )\frac{{(2\!-\!{u^2})\sin u\!-\!2u\cos u}}{{{u^3}}}\!+\!16\ln\!\!\frac{2}{{1\!+\!\cos \gamma }}\left( {\sin^2 u\!-\!{{\cos}^2} u \cos \gamma } \right) \nonumber \\
&\;\;\;\;\;\;\;\;\;+ 8(1\!-\!\cos \gamma ){\rm{ln}}\frac{{1\!+\!\cos \gamma }}{{1\!-\!\cos \gamma }}\!-\!4\frac{{\left[ {(2u\sin \frac{\gamma }{2})\cos (2u\sin \frac{\gamma }{2})\!-\!\sin (2u\sin \frac{\gamma }{2})} \right](\cos\gamma\!-\!3)\!+\!{{(2u\sin \frac{\gamma }{2})}^2}\sin (2u\sin \frac{\gamma }{2})(\cos\gamma\!-\!1)}}{{{{(2u\sin \frac{\gamma }{2})}^3}}}\nonumber  \\
&g(u)= 32\sin \frac{\gamma }{2}\frac{{\sin \frac{\gamma }{2}\sin (2u) - \sin (2u\sin \frac{\gamma }{2})}}{u} - 32{\sin ^2}\frac{\gamma }{2}\left[ {{\rm{Ci}}(2u) - {{\rm{Ci}}}(2\sin \frac{\gamma }{2}u)} \right] \nonumber  \\
&\;\;\;\;\;\;\;\;\;-\!16{\cos ^2}\frac{\gamma }{2}\begin{bmatrix} {\cos (2u)\left[ {{\rm{Ci}}(2u\!+\!2\sin \frac{\gamma }{2}u)\!+\!{\rm{Ci}}(2u\!-\!2\sin \frac{\gamma }{2}u)\!-\!2{\rm{Ci}}(2u\!-\!2{{\sin }^2}\frac{\gamma }{2}u)\!-\!2{\rm{Ci}}(2u)\!+\! 2{{\rm{Ci}}}(2{{\cos }^2}\frac{\gamma }{2}u)} \right] } \\ {\!+\!\sin (2u)\left[ {{\rm{Si}}(2u\!+\!2\sin \frac{\gamma }{2}u)\!+\!{\rm{Si}}(2u\!-\!2\sin \frac{\gamma }{2}u)\!-\!2{\rm{Si}}(2u\!-\!2{{\sin }^2}\frac{\gamma }{2}u)\!-\!2{\rm{Si}}(2u)\!+\! 2{\rm{Si}}(2{{\cos }^2}\frac{\gamma }{2}u)} \right]}\\\tag{A12} \end{bmatrix}
\end{align}
with SinIntegral ${\rm{Si}}(z)=\int_{0}^{z}\sin t /t \;dt$ and CosIntegral ${\rm{Ci}}(z)=-\int_{z}^{\infty}\cos t /t\;dt$.
\end{widetext}

\end{document}